\documentclass[useAMS,usenatbib]{mn2e}

\newcommand{\Msun}{M$_{\odot}$}
\newcommand{\kms}{km\,s$^{-1}$} 

\title[On the Star Formation History of NGC~1856]{New Constraints on the Star Formation History of the Star Cluster NGC~1856\thanks{Based on observations with the NASA/ESA {\it Hubble Space Telescope}, obtained at the Space Telescope Science Institute, which is operated by the Association of Universities for Research in Astronomy, Inc., under NASA contract NAS5-26555}}

\author[M. Correnti et al.]{Matteo Correnti$^{1}$\thanks{E-mail: correnti@stsci.edu}, Paul Goudfrooij$^{1}$, Thomas H. Puzia$^{2}$ and Selma E.\ de Mink$^{3}$ \\
$^{1}$Space Telescope Science Institute, 3700 San Martin Drive, Baltimore, MD 21218, USA\\
$^{2}$Institute of Astrophysics, Pontificia Universidad Cat\'olica de Chile, Av.\ Vicu\~{n}a Mackenna 4860, Macul 
7820436, Santiago, Chile\\
$^{3}$Astronomical Institute ``Anton Pannekoek'', University of Amsterdam, P.O.\ Box 94249, NL-1090 GE Amsterdam, The Netherlands}
\usepackage{graphicx}

\pagerange{\pageref{firstpage}--\pageref{lastpage}} \pubyear{xxxx}

\begin{document}

\date{Accepted. Received; in original form }

\maketitle

\label{firstpage}

\begin{abstract}
We use the Wide Field Camera 3 onboard the Hubble Space Telescope  to obtain
deep, high-resolution photometry of the young (age $\sim$ 300 Myr) star cluster
NGC~1856 in the Large Magellanic Cloud. We compare the observed colour-magnitude
diagram (CMD), after having applied a correction for differential reddening,
with Monte Carlo simulations of simple stellar populations (SSPs) of various
ages. We find that the main sequence turn-off (MSTO) region is wider than that
derived from the simulation of a single SSP. Using constraints based on the
distribution of stars in the MSTO region and the red clump, we find that the CMD
is best reproduced using a combination of two different SSPs with ages separated
by 80 Myr (0.30 and 0.38 Gyr, respectively). However, we can not formally
exclude that the width of the MSTO could be due to a range of stellar rotation
velocities if the efficiency of rotational mixing is higher than typically
assumed. Using a King-model fit to the surface number density profile in
conjunction with dynamical evolution models, we determine the evolution of
cluster mass and escape velocity from an age of 10 Myr to the present age,
taking into account the possible effects of primordial mass segregation.  We
find that the cluster has an escape velocity $V_{\rm esc} \simeq  17$ \kms\ at
an age of 10 Myr, and it remains high enough during a period of $\simeq$\,100
Myr to retain material ejected by slow winds of first-generation stars. Our
results are consistent with the presence of an age spread in NGC~1856, in
contradiction to the results of \citet{bassil13}. 
\end{abstract}

\begin{keywords}
galaxies: star clusters --- globular clusters: general --- Magellanic Clouds
\end{keywords}

\section{Introduction}
\label{s:intro}

Over the last $\sim$\,half a dozen years, deep colour-magitude diagrams (CMDs)
from images taken with the Advanced Camera for  Surveys (ACS) and the Wide Field
Camera 3 (WFC3) aboard the Hubble Space Telescope (HST) revealed that several
intermediate-age ($\sim$\,1\,--\,2 Gyr old) star clusters in the Magellanic
Clouds host extended main sequence turn-off (MSTO) regions
\citep{mack+08a,glat+08,milo+09,goud+09,goud+11b,goud+14,corr+14}, in some cases
accompanied by composite red clumps \citep{gira+09,rube+11}. 

A popular interpretation of these extended MSTOs (hereafter eMSTOs) is that they
are due to stars that formed at different times within the parent cluster, with
age spreads of 150\,--\,500 Myr
\citep{milo+09,gira+09,rube+10,rube+11,goud+11b,kell+12,corr+14,goud+14}.
Alternative potential causes to explain the eMSTO phenomenon presented in the
recent literature include spreads in rotation velocity among turn-off stars
\citep[][but see Girardi et al. 2011]{basdem09,yang+13}, a photometric feature
of interacting binary stars \citep{yang+11} or a combination of both
\citep{li+12}. 

An important aspect of the nature of the eMSTO phenomenon among intermediate-age
star clusters is that it is not shared by all such clusters
\citep[e.g][]{milo+09,goud+11a,corr+14}. In this context, \citet[][hereafter
G11a]{goud+11a} suggested that the key factor  in determinining whether or not a
cluster features an eMSTO is its ability to  retain material ejected by
first-generation stars  that feature relatively slow stellar outflows (the
so-called ``polluters''). Following the arguments presented in G11a, eMSTOs can
only be hosted by clusters for which the escape velocity was higher than the
wind velocity of such polluter stars at the time such stars were present in the
cluster.  

Currently, the most popular candidates for first-generation ``polluters'' are
{\it (i)\/} intermediate-mass AGB stars ($4 \la {\cal{M}}/M_{\odot} \la 8$,
hereafter IM-AGB; e.g., \citealt{danven07}, and references therein), {\it
(ii)\/} rapidly rotating massive stars (often referred to as ``FRMS''; e.g.
\citealt{decr+07}) and {\it (iii)\/} massive binary stars \citep{demink+09}.

For what concerns instead the formation scenario, the currently two favored ones
predict that the stars have been formed from or polluted by gas that is a
mixture of pristine material and material shed by such ``polluters''. In
particular, in the ``in situ star formation'' scenario \citep[see
e.g.,][]{derc+08,derc+10,conspe11}, subsequent generations of stars are formed
out of gas clouds that were polluted by winds of first generation stars to
varying extents, during a period spanning up to a few hundreds of Myr, depending
on the nature of the ``polluters''. Conversely, in the alternative ``early-disc
accretion'' scenario \citep{bast+13}, the chemically enriched material, ejected
from interacting high-mass binary systems or FRMS stars, is accreted onto the
circumstellar discs of low-mass pre-main sequence (PMS) stars that were formed
in the \emph{same generation} during the first $\simeq$ 20 Myr after the
formation of the star cluster. 

Two recent studies provided support to the predictions of the ``in situ''
scenario. \citet[][hereafter G14]{goud+14} studied in detail HST photometry of a
sample of 18 massive intermediate-age star clusters in the Magellanic Clouds
that covered a variety of masses, ages, and radii. G14 found that all the star
clusters in their sample host eMSTOs, featuring age spreads that correlate with
the clusters' escape velocity $V_{\rm esc}$, both currently and at an age of 10
Myr. 

Furthermore, \citet[][hereafter C14]{corr+14} studied 4 low-mass ($\approx 10^4$
\Msun) intermediate-age star clusters in the Large Magellanic Cloud (LMC) and
found that the two clusters that host eMSTOs have $V_{\rm esc} \ga$ 15 \kms\ out
to ages of $\sim$ 100 Myr, whereas $V_{\rm esc} \la$ 12 \kms\ at all ages for
the two clusters that do \emph{not} exhibit eMSTOs.  These results suggest that
the ``critical'' escape velocity for a star cluster to able to retain the
material ejected by the slow winds of the first generation polluters seems to be
in the approximate range of 12\,--\,15 \kms. Interestingly, these escape
velocities are consistent with wind velocities of IM-AGB stars which are in the
range 12\,--\,18 \kms\ \citep{vaswoo93,zijl+96} and with the low end of observed
velocities of ejecta of the massive binary star RY Scuti \citep[15\,--\,50
\kms,][]{smit+02,smit+07}.  

However, in the context of the multiple population scenario, one piece of the
puzzle that is still missing is the identification of an age spread (i.e.,
second-generation stars) in young massive star clusters. In fact, one of the
prediction of the ``in situ'' scenario is  that eMSTO should be observed also in
young star clusters if they have the adequate properties in terms of mass and
escape velocity. Young star clusters must be massive enough (i.e., M $\ga 10^5$
\Msun) in order to have deep enough potentials and high enough escape velocities
to retain material lost due to stellar evolution of the first generation stars
and to be able to accrete new material from the pristine gas still present in
the surroundings. Unfortunately, these strict constraints render young star
clusters in which the presence of the eMSTO phenomenon is expected and
verifiable very scarce. G11a identified the  relatively young ($\sim$ 300 Myr)
massive star cluster NGC~1856 in the LMC as a promising candidate. Following
this suggestion, \citet{bassil13} analyzed archival Wide Field \& Planetary
Camera 2  (WFPC2) images of this cluster, finding no evidence for age spreads
larger than $\sim$ 35 Myr, which was interpreted by the authors as a suggestion
that the eMSTO feature in intermediate-age clusters cannot be due to age
spreads.  However, the data analyzed by \citet{bassil13} seem to have reached
only $\simeq$\,2 mag beyond the MSTO, whose shape is vertical for the filters
used in their study (F450W and F555W). This is not deep enough to sample a
significant portion of the MS of the cluster, especially its curvature at lower
stellar masses, which is important to constrain several important parameters of
the cluster such as age, metallicity, distance, and reddening. In principle,
this could have prevented the identification of features that might point to the
presence of second-generation stars.

With this in mind, we present an analysis of new HST WFC3 photometry of
NGC~1856. These data allowed us to sample the cluster population  down to
$\simeq$\,8 mag beyond the MSTO, obtaining deep CMDs that permit a thorough
examination of the MSTO morphology and the nature of the cluster. We compare the
observed CMD with Monte Carlo simulations in order to quantify whether it can be
reproduced by one simple stellar population (SSP) or whether it is best
reproduced by a range of ages. We also study the evolution of the cluster mass
and escape velocity from an age of 10 Myr to its current age, to verify whether
the cluster has the right properties to form and retain a second generation of
stars. This analysis allows us to reveal new findings on the star formation
history of the cluster NGC~1856 in the context of the multiple population
scenario.  
 
The remainder of the paper is organized as follows: Section~\ref{s:obs} presents
the observations and data reduction. In Section~\ref{s:cmd} we present the
observed CMD, we describe the technique applied to correct the CMD due to the
presence of differential reddening effects and we derive the best-fit
isochrones. In Section~\ref{s:mcsim} we compare the observed CMD with Monte
Carlo simulations, both using one SSP and a combination of two SSPs with
different ages. We derive pseudo-age distributions for the observed and
simulated CMDs and compare them. In Section~\ref{s:excess} we investigate the
presence of ongoing star formation in the cluster, looking for PMS stars of the
second generation. In Section~\ref{s:rotation} we test the prediction of the
stellar rotation scenario, while  Section~\ref{s:dynamics} presents the physical
and dynamical properties of the  cluster, deriving the evolution of the cluster
escape velocity as a function of age. Finally, in Section~\ref{s:conclusion}, we
present and discuss our main results.

\section{Observations and Data Reduction}
\label{s:obs}

\begin{figure}
\includegraphics[width=1\columnwidth]{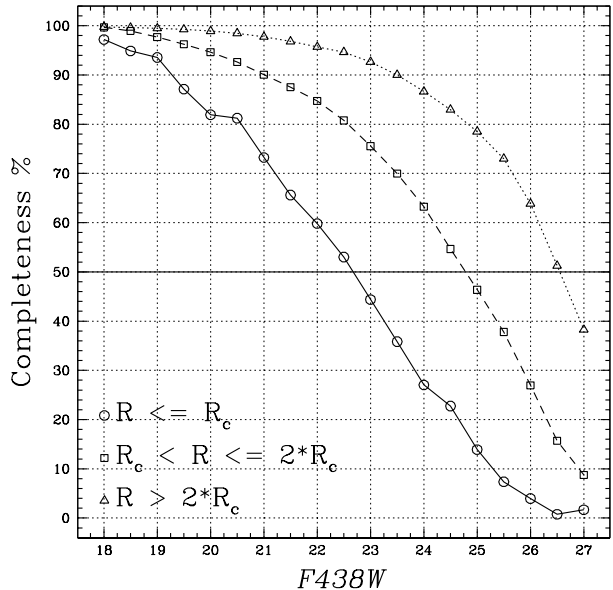}
\caption{Completeness fraction as function of magnitude F438W and distance from
the cluster center. The open circles and solid lines represent data within the
core radius $r_c$, based on the \citet{king62} model fit derived as described in
Section~\ref{s:kingmodel}; open squares and dashed line represent data in the
range between $r_c$ and 2 $\times r_c$ and the open triangles and dotted lines
represent data outside 2 $\times r_c$.}
\label{f:completeness}
\end{figure}

NGC~1856 was observed with HST on 2013 November 12 using the UVIS channel of the
WFC3 as part of the HST program 13011 (PI: T. H.\ Puzia). The cluster was
centered on one of the two CCD chips of the WFC3 UVIS camera, so that the
observations cover enough radial extent to study variations within cluster
radius and to avoid the loss of the central region of the cluster due to the CCD
chip gap. The cluster was observed in four filters, namely F438W, F555W, F658N,
and F814W. Two long exposures were taken in each of the four filters: their
exposure times were 430 s (F438W), 350 s (F555W), 450 s (F814W) and 940 s
(F658N). In addition, we took three short exposures in the F438W, F656N and
F814W filters (185 s, 735 s and 51 s, respectively), to avoid saturation of the
brightest RGB and AGB stars. The two long exposures in each filter were
spatially offset from each other by 2\farcs401 in a direction +85\fdg76 with
respect to the positive X-axis of the detector, in order to move across the gap
between the two CCD chips, as well to simplify the identification and removal of
hot pixels and cosmic rays. In addition to the WFC3/UVIS observations, we used
the Wide Field Camera (WFC) of ACS in parallel to obtain images $\approx 6'$
from the cluster center. These ACS images have been taken with the F435W, F656N
and F814W filters and provide valuable information of the stellar content and
star formation history in the underlying LMC field, permitting us to establish
in detail the field star contamination fraction in each region of the CMDs.

To reduce the images we followed the method described in \citet{kali+12}.
Briefly, we started from the \emph{flt} files provided by the HST pipeline,
which constitute the bias-corrected, dark-subtracted and flat-fielded images.
After correcting the \emph{flt} files for charge transfer inefficiency using the
dedicated CTE correction
software\footnote{http://www.stsci.edu/hst/wfc3/tools/cte\_tools}, we generated
distortion-free images using MultiDrizzle \citep{fruchook02} and we calculated  
the transformation between the individually drizzled images in each filter,
linking them to a reference frame (i.e., the first exposure). Through these
transformations we obtain an alignment of the individual images to better than
0.02 pixels. After flagging and rejecting bad pixels and cosmic rays from the
input images, we created a final image for each filter, combining the input
undistorted and aligned frames. The final stacked images were generated at the
native resolution of the WFC3/UVIS and ACS/WFC (i.e., 0\farcs040 pixel$^{-1}$
and 0\farcs049 pixel$^{-1}$, respectively).  

\begin{table}
\begin{center}
\begin{tabular}{ccccc}
\hline
\hline
R & F438W & F555W & F658N & F814W \\
 (1) & (2) & (3) & (4) & (5)\\
\hline
$R < R_c$ & 22.68 & 22.50 & 21.56 & 22.06\\
$R_c < R < 2\cdot R_c$ & 24.78 & 24.52 & 22.33 & 23.88\\
$R > 2\cdot R_c$ & 26.55 & 26.48 & 22.60 & 25.76\\
\hline
\end{tabular}
\caption{50\% completeness limit for each bands in three different intervals in radius. (1): Radius interval (2-5): magnitudes in each band corresponding to 50\% completeness fraction.}
\label{t:compl}
\end{center}
\end{table}

To perform the stellar photometry, we used the stand-alone versions of the
DAOPHOT-II and ALLSTAR point spread function (PSF) fitting programs
\citep{stet87,stet94} on the stacked images. To obtain the final catalog we
first performed the aperture photometry on all the sources that are at least
3$\sigma$ above the local sky, then we derived a PSF from $\sim$ 1000 bright
isolated stars in the field, and finally we applied the retrieved PSF to all the
sources detected through the aperture photometry. We retained in the final
catalogs only the sources that were iteratively matched between the two images
and we cleaned them eliminating background galaxies and spurious detections by
means of $\chi^2$ and sharpness cuts from the PSF fitting.   

Photometric calibration has been performed using a sample of bright isolated
stars to transform the instrumental PSF-fitted magnitudes to a fixed aperture of
10 pixels (0\farcs40 for WFC3/UVIS, 0\farcs49 for ACS/WFC). We then transformed
the magnitudes into the VEGAMAG system by adopting the relevant synthetic zero
points for the WFC3/UVIS and ACS/WFC filters. Positions and magnitudes, with the associated errors, for the first ten objects in our final catalog are reported in Table~\ref{t:phot} in the Appendix~\ref{s:photometry}.

We characterized the completeness as well as the photometric error distribution
of the final photometry by performing artificial star tests, using the standard
technique of adding artificial stars to the images and running them through the
photometric routines that were applied to the drizzled images using identical
criteria. We added to each image a total of nearly 700000 artificial stars. To
not induce incompleteness due to crowding in the tests themselves, the fraction
of stars injected into a given individual image was set to be $\sim$ 5\% of the
total number of stars in the final catalogs. The overall distribution of the
inserted artificial stars was chosen to follow that of the stars in the image.
We distributed the artificial stars in magnitude in order to reproduce a
luminosity function similar to the observed one and with a colour distribution
that span the full colour ranges found in the CMDs. After inserting the
artificial stars, we applied to each image the photometry procedures described
above. The stars were recovered blindly and automatically cross-matched to the
input starlist containing actual positions and fluxes; they were considered
recovered if the input and output magnitudes agreed to within 0.75 mag in both
filters. Finally, we assigned a completeness fraction to each individual star in
a given CMD as a function of its magnitude and distance from the cluster center.
The completeness fraction of stars as a function of the magnitude F438W and
distance from the cluster center, in three different intervals in radius, are
shown in Figure~\ref{f:completeness}. The magnitudes corresponding to the 50\% completeness fraction in each band, for the same intervals, are reported in Table~\ref{t:compl}. 

\section{Colour-Magnitude Diagram Analysis}
\label{s:cmd}

\begin{figure}
\includegraphics[width=1\columnwidth]{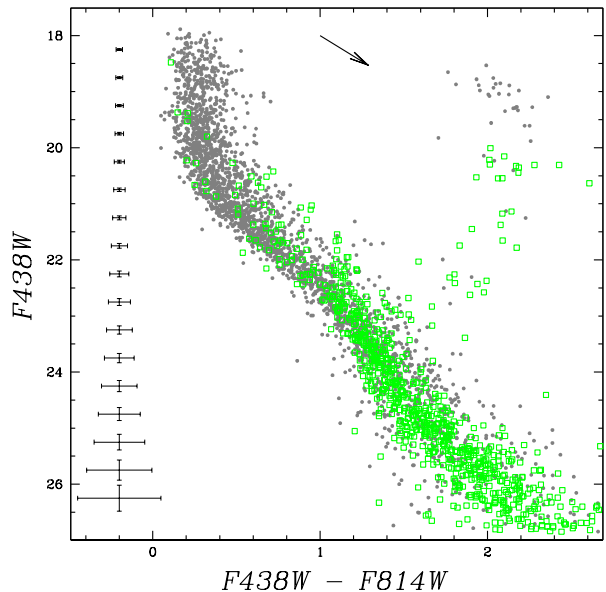}
\caption{Observed F438W vs.\ F438W $-$ F814W CMD for all the stars inside the
core radius, based on the \citet{king62} model fit derived as described in
Section~\ref{s:kingmodel}. Contamination from the underlying LMC field
population has been derived from a region near the corner of the image, with the
same surface area adopted for the cluster stars, and superposed on the cluster
CMD (green open squares). Magnitude and colour errors, derived using the
photometric distribution of our artificial stars within the core radius are
shown in the left side of the CMD. The reddening vector is also shown for $A_V =
0.4$.}
\label{f:cmdobs}
\end{figure}

\subsection{A wide main sequence turn-off region}
\label{s:eMSTO}

Figure~\ref{f:cmdobs} shows the F438W $-$ F814W vs.\ F438W CMD of NGC~1856, 
plotting only the stars within the core radius, based on the \citet{king62}
model fit derived as described in Section~\ref{s:kingmodel}.  The observed CMD
presents some interesting features. While the lower part of the MS (i.e., below
the ``kink'' at F438W $\simeq$~21.2) is quite narrow and well defined, the MSTO
region (with F438W $\la$~19.5), appears relatively wide when compared with the
photometric errors. 

To determine whether the observed broadening of the MSTO region is a ``real''
feature in the CMD, we first test whether it could be caused by the following
effects: contamination by the underlying field population, poor photometry
(i.e.; large photometric uncertainties), the presence of differential reddening,
and/or a significant binary fraction.  

To assess the level of contamination of the underlying LMC field population, we
selected a region near the corner of the image with the same surface area
adopted for the cluster stars. Stars located in this background region have been
superposed on the cluster's CMD (shown as green open squares in
Figure~\ref{f:cmdobs}). The contamination is mainly confined to the lower
(faint) part of the MS and does not affect the MSTO and the Red Clump (RC) in
any significant fashion; thus we can conclude that underlying LMC field stars do
not cause the observed shape of the MSTO or RC regions.  

Photometric uncertainties have been derived from the artificial stars test.
Magnitude and color errors are shown in Figure~\ref{f:cmdobs}; photometric
errors at the MSTO level are between 0.02 and 0.04 mag, far too small to account
for the broadening of the MSTO. 

\begin{figure}
\includegraphics[width=1\columnwidth]{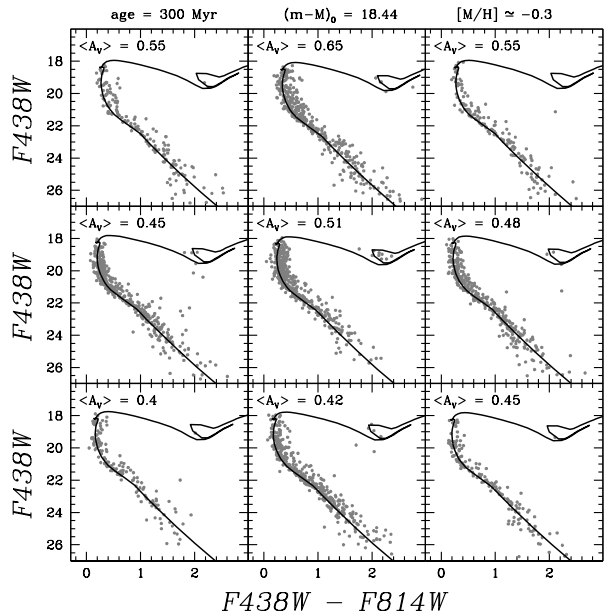}
\caption{CMDs for the 9 subregions in which we divided the cluster field,
derived as described in Section~\ref{s:cmd}. Best-fit isochrones (black lines)
from \citet{mari+08} are superposed in each CMD, along with the derived visual
extinction $A_V$. The top legend shows the age, distance modulus $(m-M)_0$ and
metallicity adopted for all the isochrones.}
\label{f:cmdreg}
\end{figure}

\subsection{The impacts of differential reddening and binary fractions}
\label{s:difred}

\begin{figure}
\hspace{-0.75truecm}
\includegraphics[width=9truecm]{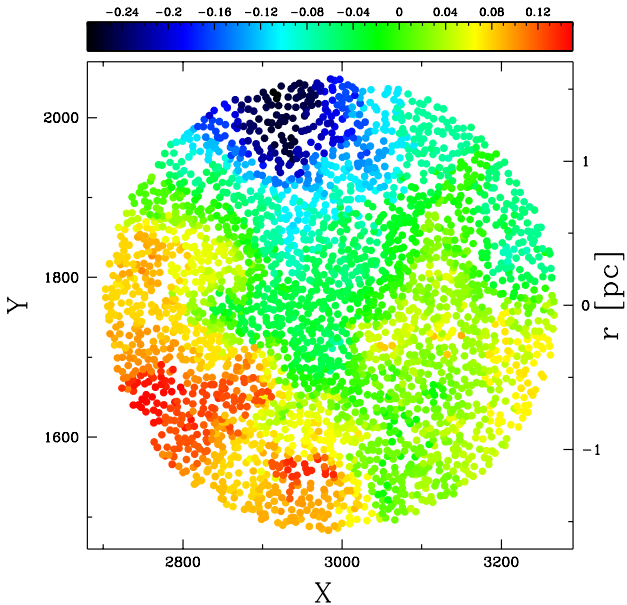}
\caption{Spatial distribution of the differential reddening in NGC~1856. The
color of each star represents the final $\Delta E(B-I)$ derived from
the differential reddening correction inside the core radius region. The color coding is shown at the top.}
\label{f:map}
\end{figure}

To check for the presence of differential reddening in the cluster field we adopted the following approach. First of all, we selected a square region that circumscribes the area defined by the
core radius (i.e.; the length of the side of the square is equal to twice the
core radius), we divided this region into 9 equal-sized squares, and we compared
the CMDs derived in each of them. The number of equal-sized region has been arbitrarily choosen in order to have a good sampling of the different region mantaining a statistically significant number of objects in each field. This approach has been adopted only to check if differential reddening effects are present in the cluster field and to have a rough estimate of the reddening variation. Figure~\ref{f:cmdreg} shows the CMDs for the
9 subregions; to yield a preliminary hint on the amount of differential
reddening, we superimposed on each CMD an isochrone from \citet{mari+08} for
which age, distance modulus and metallicity are fixed (see top legend of
Figure~\ref{f:cmdreg}), while the  visual extinction $A_V$ is left free to vary.
Figure~\ref{f:cmdreg} clearly shows that the best fit is achieved in each CMD
using a different value of the visual extinction $A_V$ (derived values of $A_V$
are reported in each field), confirming that a differential reddening is present
in the cluster field, with variations of the order of $\sim \pm$ 0.15 mag with
respect to the mean reddening value. 

Once verified that the differential reddening is truly present in the cluster field, we corrected each star in the CMD using the
following approach \cite[for a more detailed description, see][]{milo+12}.
Briefly, we first defined a photometric reference frame in which the X axis is
parallel to the reddening line. In this reference system, it is much easier to
determine reddening difference rather than in the original colour-magnitude
plane, where the reddening vector is an oblique line. To do this, we arbitrarily
defined an origin {\it O}, then translated the CMD such that the origin of the
new reference frame corresponds to {\it O} and then we rotated the CMD
counterclockwise by an angle defined by the equation: 
\begin{equation}
\theta = \arctan \left( \frac{A_{F438W}}{A_{F438W} - A_{F814W}} \right)
\end{equation}
where $A_{F438W}$ and $A_{F814W}$ are the appropriate extinction coefficients for the UVIS WFC3 filters. Using \citet{card+89} and \citet{odon94} extinction curve and adopting $R_V = 3.1$, their values are $A_{F438W}$ = 1.331$\cdot A_V$ and $A_{F814W} $ = 0.610$\cdot A_V$, respectively. 
At this point, we generated a fiducial line using only MS stars. We divided the
sample into bins of fixed ``magnitude'' and we calculated the medians in the X-
and Y-axis. The use of the median allows us to minimize the influence of
outliers such as binary stars or stars with poor photometry left in the sample
after the selection. Among the MS stars, we selected a subsample located in the
region where the reddening line defines a wide angle with the fiducial line, so
that the shift in colour can be safely interpreted as an effect of the
differential reddening. For this reason, we limited our sample to the central
portion of the cluster MS, excluding the upper part near the MSTO and the lower
portion, fainter than the magnitude at which the MS starts to bend in a
direction parallel to the reddening line. For each selected star, we calculated
the distance from the fiducial line along the reddening direction (i.e. $\Delta$
X) and we used these stars as reference stars to estimate the reddening
variations associated to each star in the CMD. Finally, to correct them, we
selected the nearest 30 reference stars and we calculate the median $\Delta$ X
that is assumed as the reddening correction for that particular star (to derive
the differential reddening suffered by the reference stars, we excluded that
star in the computation of the median $\Delta$ X). We note that the median
correlation length introduced with this approach is of the order of $\simeq$\,20
px (corresponding to 0.2 pc). After the median $\Delta$ X values have been
subtracted from the X-axis value of each star in the rotated CMD, we obtained an
improved diagram that we used to derive a more accurate selection of MS
reference stars and a more precise fiducial line. We then applied again the
described procedure, iterating the process until it converges (in this case, a
couple of iterations were sufficient). At this point, the corrected X and Y
``magnitude'' were converted back to the F438W and F814W magnitudes.  

Figure~\ref{f:map} shows the spatial distribution of the differential reddening
inside the core radius of NGC~1856; each star is reported with a different
color, depending on the final $\Delta E({\rm F438W}-{\rm F814W})$ applied to
correct it, derived  with the method described above. In this context, we
observe that the CMD in the central panel of Figure~\ref{f:cmdreg}, which
represents the innermost region of the cluster, shows a quite wide MSTO,
compared to the other panels. 

Conversely, in the corresponding spatial region from which the CMD is drawn
(i.e., the central part in the map of Figure~\ref{f:map}) the derived reddening
is quite uniform, with $\Delta E({\rm F438W}-{\rm F814W}) \simeq$ 0.04 
(corresponding to $\Delta A_V \simeq$ 0.055).   Moreover, it is worth to note
that the reddening correction reaches its highest accuracy in this inner region,
due to the high surface number density  of stars (i.e., smaller distance of the
star to be corrected from the reference stars used to derive $\Delta A_V$, see
also Section~\ref{s:ssp1}). 

\begin{figure*}
\includegraphics[width=11cm]{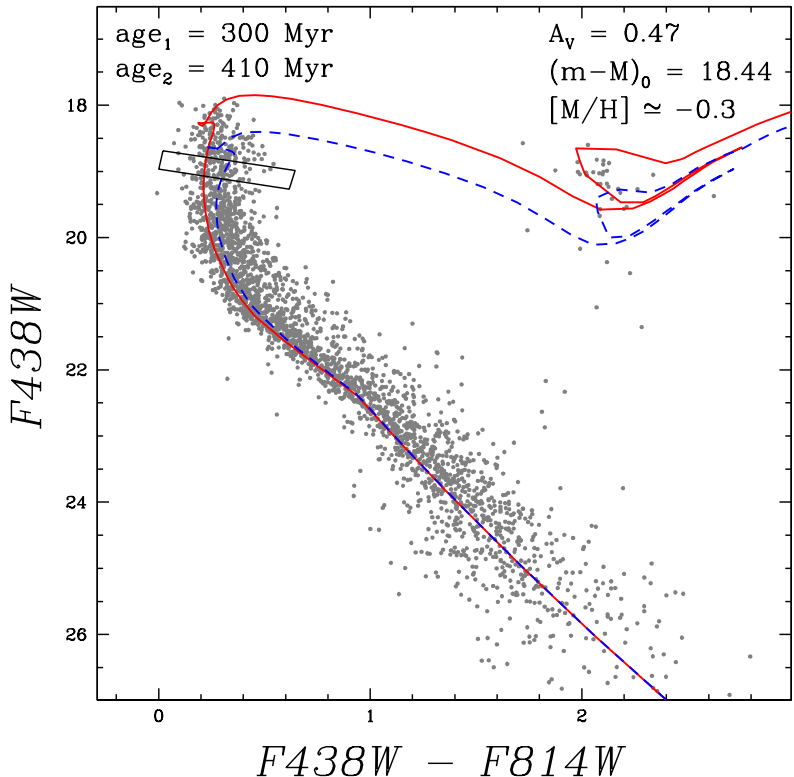}
\caption{Differential-reddening-corrected F438W - F814W versus F438W CMD for
 NGC~1856. Best-fit isochrones from \citet{mari+08}, for the minimum (red solid
 line) and maximum (blue dashed line) ages (300 and 410 Myr, respectively) that
 can be accounted for by the data are superposed on the cluster CMD, along with
 the derived distance modulus $(m-M)_0$ and visual extinction $A_V$. The
 parallelogram box used to select MSTO stars in order to derive the pseudo-age
 distribution, as described in Section~\ref{s:mcsim}, is also shown. }
\label{f:cmdcorr}
\end{figure*}

We acknowledge that our differential reddening correction might be somewhat
ineffective for a \emph{small} fraction of stars, depending on their location in
the field,  e.g., in the outer regions, where the distance between stars is
generally larger than in the inner regions. However, the number of stars in such
regions is by nature relatively low (see, e.g.,  the CMDs in the corner panels
of Figure~\ref{f:cmdreg}). Thus, only a small fraction of MSTO stars can suffer
from a less accurate correction, leaving the global morphology of the MSTO
region unaltered. 

\begin{figure}
\includegraphics[width=1\columnwidth]{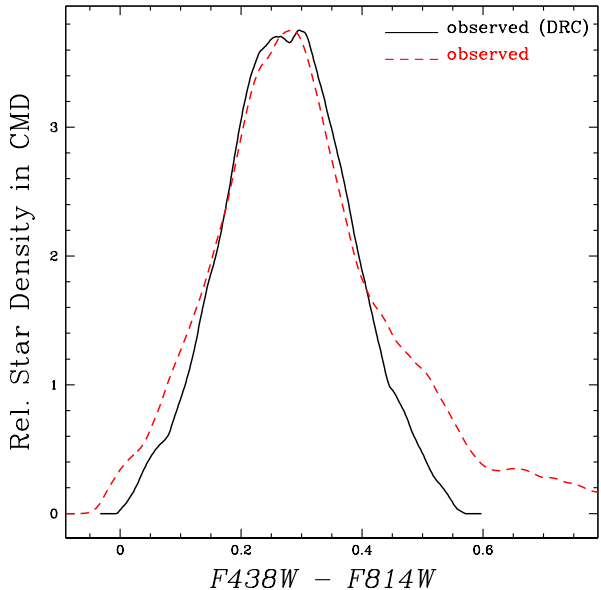}
\caption{Comparison between the colour distributions obtained from the
  differential-reddening-corrected CMD (black solid line) and the
  observed CMD (red dashed line).} 
\label{f:cfr_obs}
\end{figure}

The differential-reddening-corrected (DRC) CMD is shown in
Figure~\ref{f:cmdcorr}. The correction caused the lower part of the MS to be
even more defined and narrow than it was in the observed CMD. Conversely, the
MSTO region and upper MS is still fairly wide, although it has narrowed down
slightly.  This is illustrated in more detail in Figure~\ref{f:cfr_obs} which
shows the distribution of F438W\,$-$\,F814W colours in the magnitude range 
$18.7 \leq {\rm F438W} \leq 19.2$\footnote{This magnitude range was selected in
this context because it reflects the part of the upper MS for which the binary
sequence is merged in with the single star sequence (see Fig.~\ref{f:cfr1}).
Hence, our result is robust against any given binary fraction.} mag for the
observed CMD and that obtained for the DRC CMD (red dashed and black solid
lines, respectively). Note that the global shape and width of the two
distributions is very  similar, except for the fact that the observed one has
more extended outer  wings than the one derived from the DRC CMD. These extended
wings are caused by stars that are located in the region where the differential
reddening effects are stronger and then have been shifted from their original
position in the CMD. The fact that the shape of the two distributions is very
similar within their FWHM values suggests that the observed broadening in the
MSTO region is an intrinsic feature, rather than due to differential reddening. 

Given the above, we judge it very unlikely that differential
reddening effects or binary stars can explain the observed broadening of the MSTO region.   

\subsection{Isochrone fitting}
\label{s:iso}

Isochrone fitting was done following the methods described in detail in
\citet[][hereafter G11b]{goud+11b} for the case of star clusters with ages that
are too young to have a well-developed red giant branch. Briefly, age and
metallicity were derived using the observed differences in magnitude and colour
between the MSTO and the RC; we selected all the isochrones for which the values
of those parameters lie within $2\sigma$ of the  uncertainties of those
parameters as derived from the CMDs. For the set of isochrones that satisfied
our selection criteria (5\,--\,10 isochrones), we found the best-fit values  for
distance modulus and reddening by means of a least-square fitting program to the
magnitudes and colors of the MSTO and RC. Finally, we overplotted the isochrones
onto the CMDs and selected the best-fitting ones by means of a visual
examination. 

In this context, we superposed on the cluster CMD two isochrones from
\citet{mari+08} of different ages (300 Myr and 410 Myr, respectively), shown  in
Figure~\ref{f:cmdcorr} with different colours and line styles (red solid line
for the younger isochrone and blue dashed line for the older one, respectively).
The isochrones ages were chosen to match the minimum and maximum age that can be
accounted for by the data and isochrone fitting performed as described above. 

Taking the results shown in Sections \ref{s:eMSTO}\,--\,\ref{s:iso} at face
value,  it seems that the morphology of the MSTO is not very well reproduced by
a SSP, and that a spread in age of the order of $\sim$ 100 Myr may constitute a
better fit to the data.  This is further addressed below (Section~\ref{s:ssp1}
and Section~\ref{s:ssp2}).   

\section{Monte Carlo Simulations}
\label{s:mcsim}

To further determine whether or not a single population can reproduce the
observed CMD, we conducted Monte Carlo simulations of a synthetic cluster with
the properties implied by the isochrone fitting (see G11a and G14 for a detailed
description of the method adopted to produce these simulations). Briefly, to
simulate a SSP with a given age and chemical composition, we populated an
isochrone with stars randomly drawn using a Salpeter mass function and
normalized to the observed (completeness-corrected) number of stars. To a
fraction of these sample stars, we added a component of unresolved binary stars
derived from the same mass function, using a flat distribution of
primary-to-secondary mass ratios. The binary fraction was estimated by
using as a template the part of the observed MS between the MSTO region and the
TO of the background population in the magnitude range $22.5 \la F438W \la
19.5$ (see Fig.~\ref{f:cmdobs}). We derived that the value of the binary
fraction that best fit the data is of $\simeq$ 25\%. We estimated that the
internal systematic uncertainty in the binary fraction is $\simeq$\,5\%; for the
purposes of this work, the results do not change significantly within 10\% of
the binary fraction.  Finally, we added photometric errors that were derived
using our artificial star tests. We derived three different sets of synthetic
CMDs: one with a single SSP of age 300 Myr, and two combining two SSPs of
different ages (one with SSPs with ages of 300 Myr and 380 Myr, and the other
with ages of 300 Myr and 410 Myr, respectively). In the last two cases, we
derived a set of simulations in which we changed the ratio of the number of
stars in the two populations, from 10\% to 90\% of stars in the younger SSP,
with a 5\% step increase in each run.  

To compare in detail the observed MSTO region with the simulated ones, we
created ``pseudo-age'' distributions (see G11a for a detailed description).
Briefly, pseudo-age distributions are derived by constructing a  parallelogram
across the MSTO, with one axis approximately parallel to the isochrones and the
other approximately parallel to them (the adopted parallelogram is shown in
Figure~\ref{f:cmdcorr}). The (F438W\,$-$\,F814W, F438W) coordinates of the stars
within the parallelogram are then transformed into the reference coordinate
frame defined by the two axis of the parallelogram and the same procedure is
applied to the isochrones tables to set the age scale along this vector. The
pseudo-age distributions are calculated using the non-parametric
Epanechnikov-kernel density function \citep{silv86}, in order to
avoid possible biases that can arise if fixed bin widths are used. Finally, a
polynomial least-squares fit between age and the coordinate in the direction
perpendicular to the isochrones yields the final pseudo-age distributions.  The
described procedure is applied both to the observed and the simulated CMD. In
the following sections, we describe the results obtained from the comparison of
the observed and simulated pseudo-age distributions, for the synthetic CMDs
obtained with a single SSP and for those obtained from the combination of two
SSPs of different ages. 

\subsection{Comparison with the synthetic CMD of a SSP}
\label{s:ssp1}

\begin{figure}
\includegraphics[width=1\columnwidth]{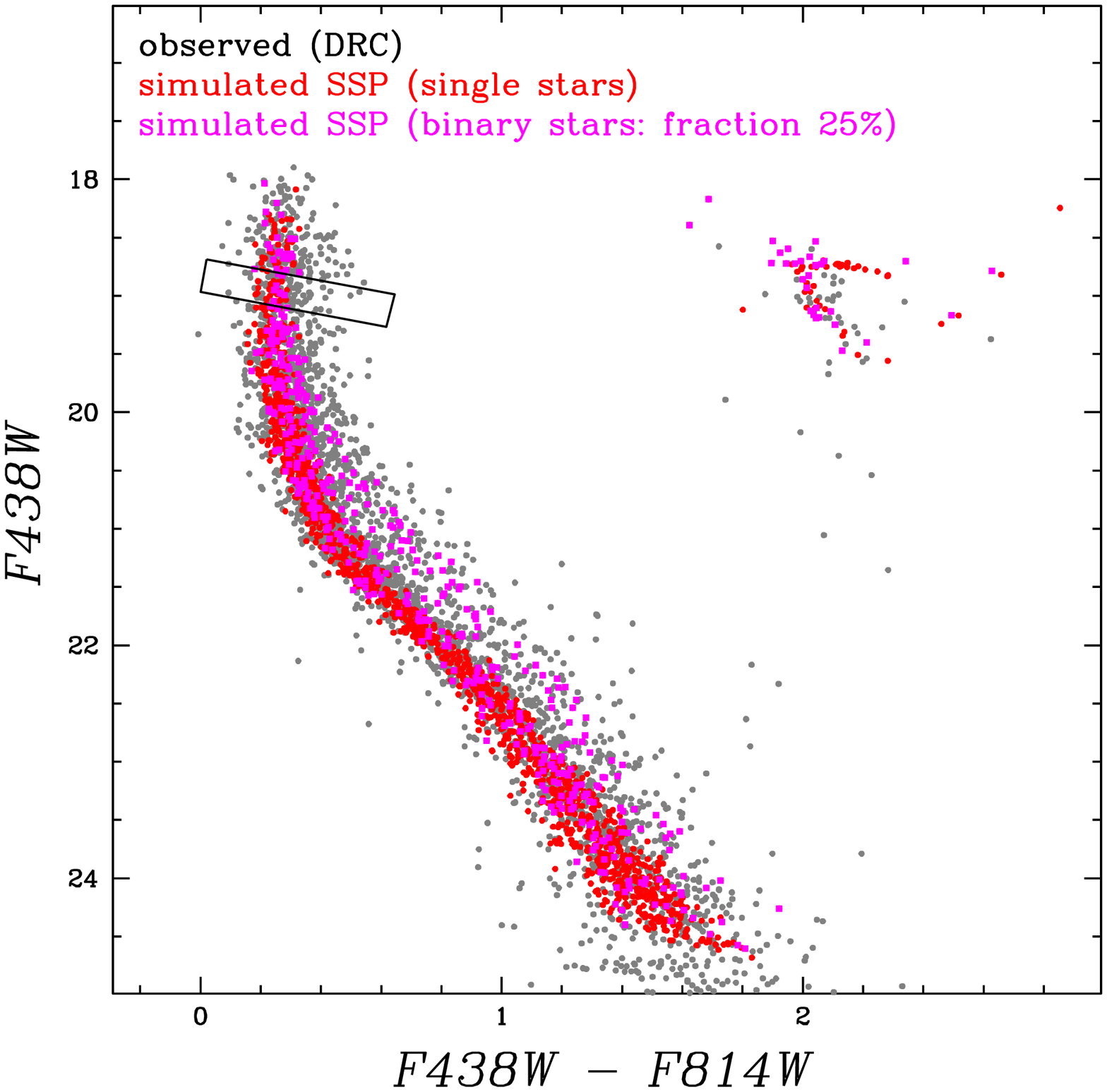}\\
\includegraphics[width=1\columnwidth]{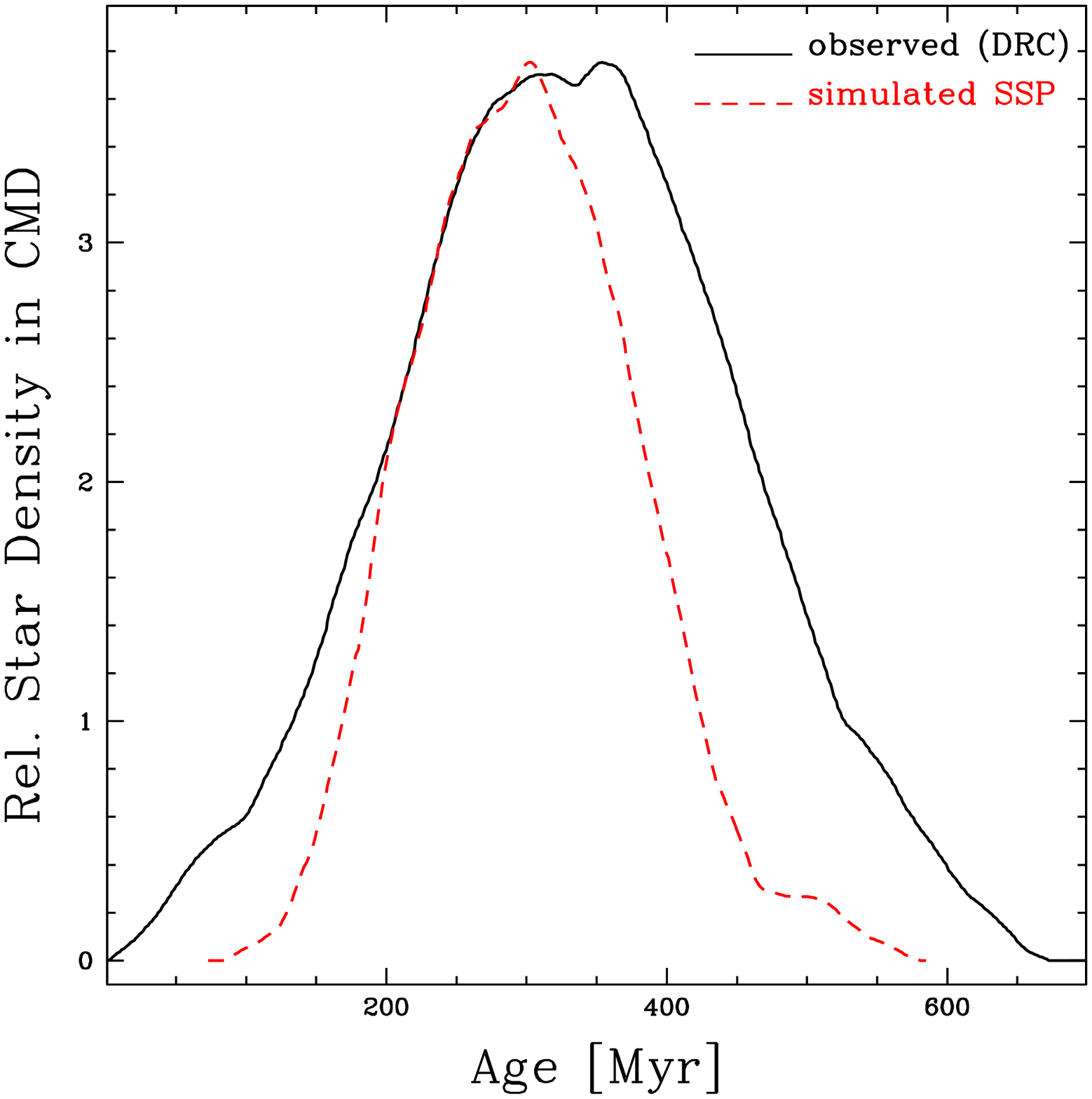}
\caption{Top panel: comparison between DRC (gray dots) and simulated  CMDs (red
 circles, single stars, magenta square, binary stars, respectively). The
 simulated CMD is obtained from Monte Carlo simulation of a SSP with an age of
 300 Myr. The parallelogram box used to select MSTO stars and the fraction of
 the adopted binary stars are also reported. Bottom panel: Pseudo-age
 distribution for the MSTO region of the DRC (black solid line) and simulated
 (red dashed line) CMDs.}
\label{f:cfr1}
\end{figure}

The top panel of Figure~\ref{f:cfr1} shows the comparison between the observed
and simulated CMD, the latter obtained from a single SSP with an age of 300 Myr.
Overall, the SSP simulation reproduces the CMD features quite well  in the
fainter portion of the cluster MS (i.e.; for F438W $\ga$ 20.5 mag). Conversely,
we note that the MSTO region\footnote{i.e., the part of the MS with F438W $\la$
19.2 mag, where the binary sequence has joined the single star sequence in
F438W\,$-$\,F814W} of the simulated SSP does not reproduce the observed MSTO
region very well, in that the latter is wider than the former.   Furthermore,
the faint end of the RC also does not seem to be reproduced well by the
simulated SSP in that the simulated luminosity function of RC stars  peaks more
strongly at bright magnitudes than the observed one.  Note that this is
consistent with a fraction of stars in NGC~1856 having ``older'' ages than that
of the best-fit SSP (see for reference the shape and location of the RC for the
older isochrone in Figure~\ref{f:cmdcorr}).   This difference is even more
evident in a comparison of the observed and simulated pseudo-age distributions
(see bottom panel of Figure~\ref{f:cfr1}). Indeed, the observed pseudo-age
distribution (shown as a black solid line in Figure~\ref{f:cfr1}) reaches
significantly older ages than the simulated one (red dashed line in
Figure~\ref{f:cfr1}), while the two distributions are very similar to each other
in the left (``young'') half of the respective profiles. 

To quantify the difference between the pseudo-age distribution of the cluster
data and that of the SSP simulation in term of {\it intrinsic} MSTO width of the
cluster, we measured the widths of the two sets of distributions at 50\% of
their maximum values (hereafter called FWHM), using quadratic interpolation. The
intrinsic pseudo-age range of the cluster is then estimated by subtracting the
simulation width in quadrature: 
\begin{equation}
{\it FWHM}_{\rm MSTO} = ({\it FWHM}^{2}_{\rm obs} - {\it FWHM}^{2}_{\rm SSP})^{1/2}
\label{eq:fwhm}
\end{equation}
where the ``obs'' subscript indicates a measurement on the DRC CMD and the
``SSP'' subscript indicates a measurement on the simulated CMD for the SSP. 
With ${\it FWHM}_{\rm obs}$ = 269 Myr and ${\it FWHM}_{\rm SSP}$ = 198 Myr, we
obtain ${\it FWHM}_{MSTO} \approx$ 182 Myr, which is similar to the value of
${\it FWHM}_{\rm SSP}$. This suggests that NGC~1856 may host two SSPs separated
in age by about ${\it FWHM}_{MSTO}/2  \sim$\,90 Myr (see Sect.~\ref{s:ssp2}
below). 

In this context, we adopted equation~\ref{eq:fwhm} also to derive the value of
reddening variation necessary to reproduce the observed pseudo-age distributions
assuming that a cluster is formed by a SSP. Instead of age, we derived the FWHM
of the observed and simulated pseudo-age distributions as a function of F438W -
F814W; using the appropriate extinction coefficient for the UVIS WFC3 filters we
transformed the ``$\Delta$ color'' in $\Delta A_V$ obtaining the following
value: $\Delta A_V$ = 0.27. The derived $\Delta A_V$ is significantly larger
than the reddening variations observed in the spatial differential reddening map
in the innermost region of the cluster. Moreover, we note that due to the
uniform shape of the observed pseudo-age distribution, this large reddening
should affect a significant number of stars, making the possibility that the
reddening can mimic the presence of a multiple population even less plausible.
Therefore, taking these results at face value, they seem to suggest that a SSP
is not able to reproduce the observed morphology of the MSTO region.

\begin{figure}
\includegraphics[width=1\columnwidth]{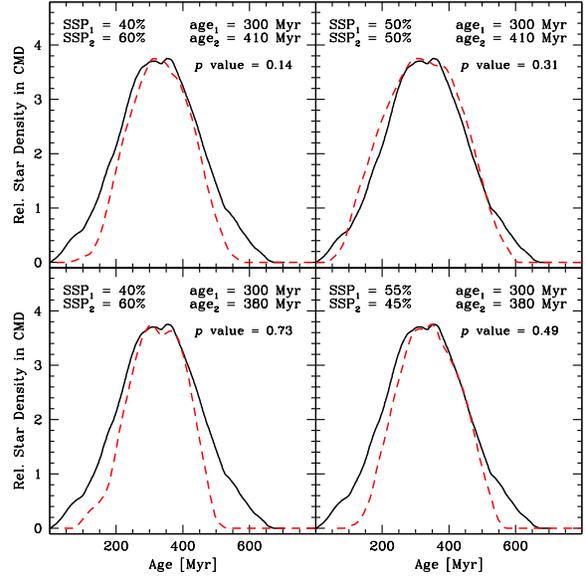}
\caption{Comparison between the observed pseudo-age distribution (from the DRC
  CMD, black solid line) and the ones obtained from the combination of two SSPs
  with different ages (300 Myr and 410 Myr, top panels; 300 Myr and 380 Myr,
  bottom panels) shown as red dashed lines. In each panel we report the adopted
  ages of the two SSPs and the number ratio between the younger SSP and the old
  one. Two-tail {\it p} values obtained for the K-S test are also shown.} 
\label{f:cfr2}
\end{figure}

\subsection{Comparison with a synthetic CMD of two SSPs of different age}
\label{s:ssp2}

\begin{figure*}
\includegraphics[width=1\columnwidth]{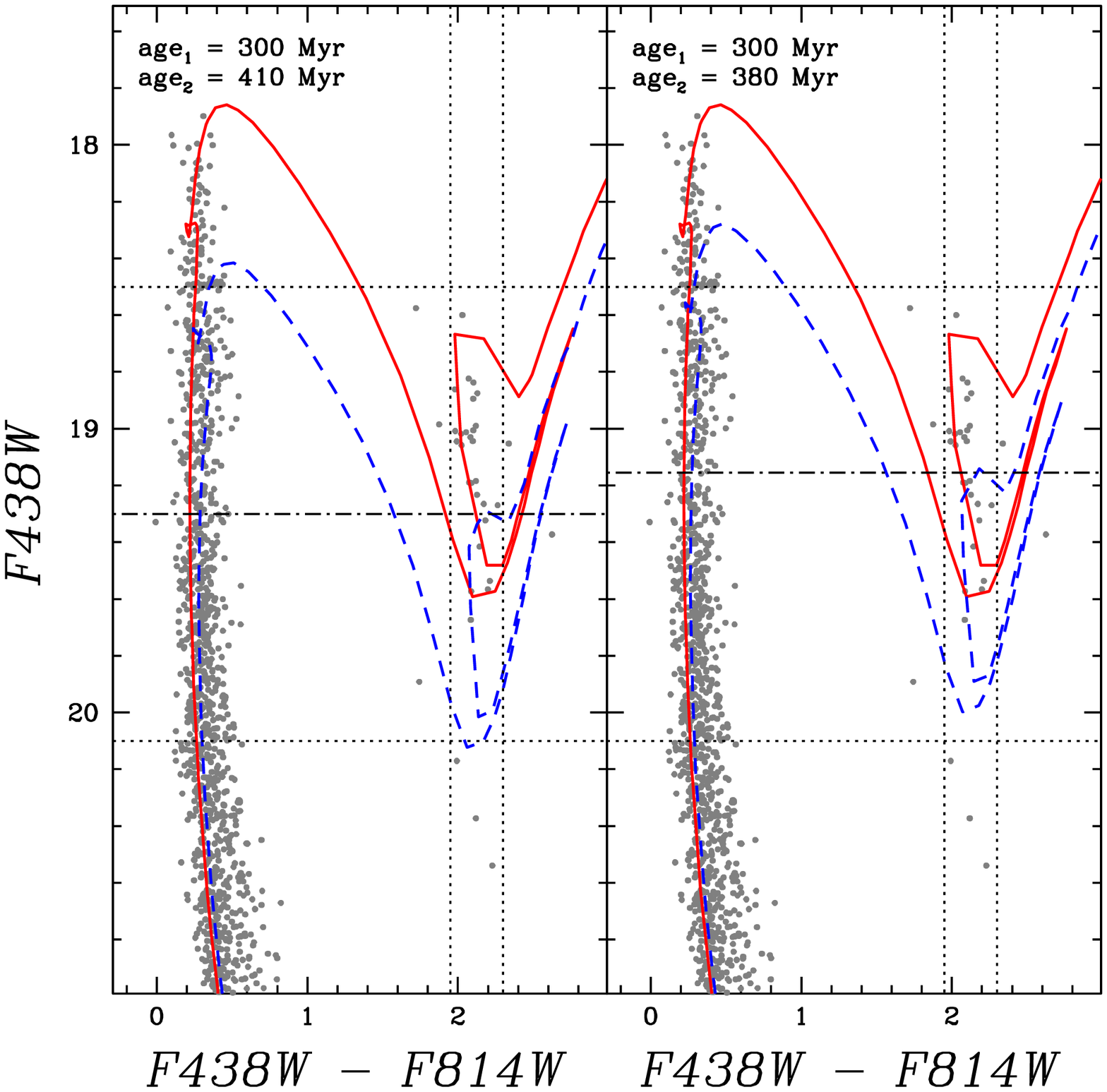}
\includegraphics[width=1\columnwidth]{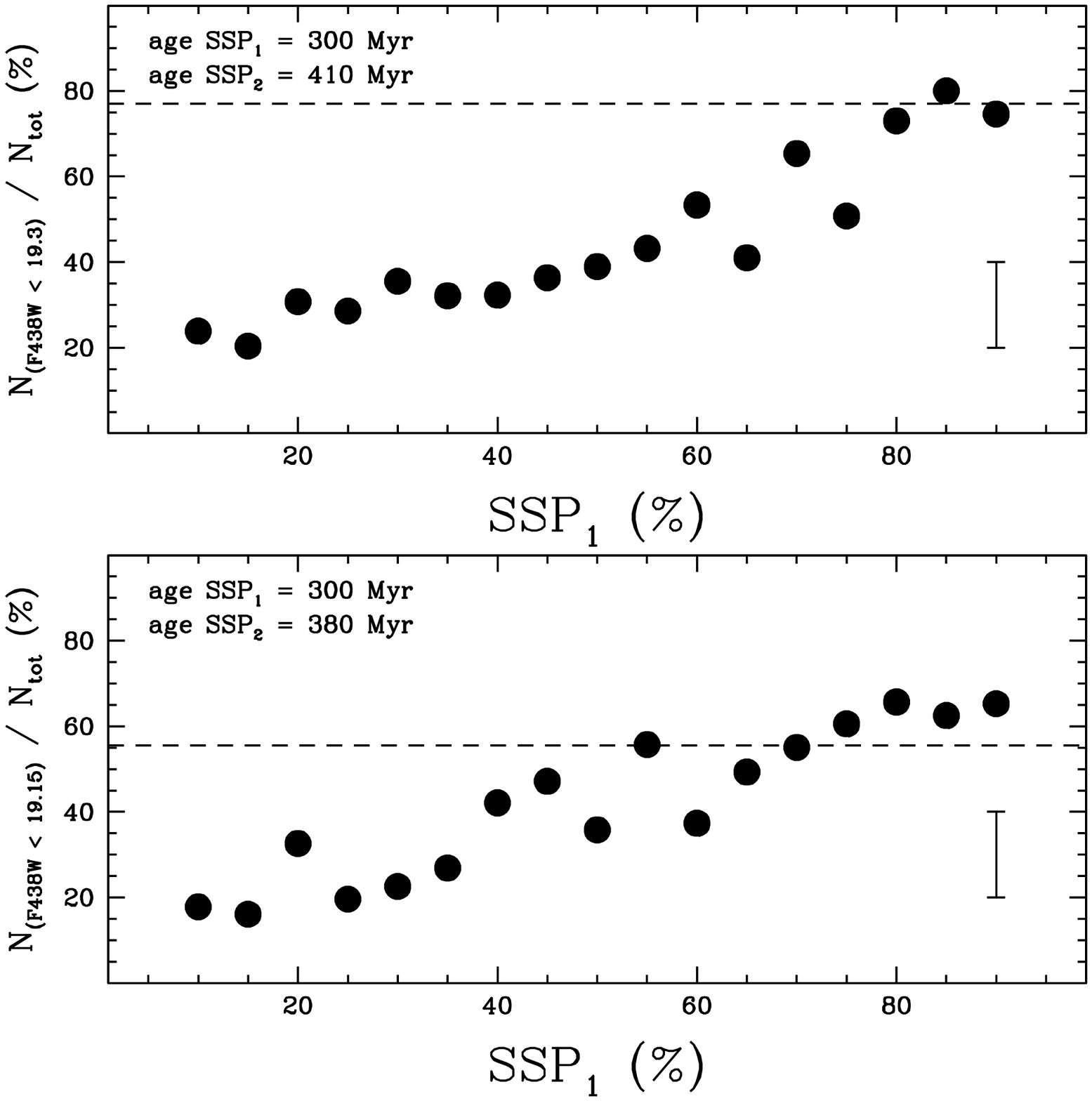}
\caption{Left panels: DRC CMDs zoomed in the upper portion of the MS with
  superimposed isochrones (red solid and blue dashed lines for the young and the
  old isochrones, respectively) from \citet{mari+08} along with the adopted ages
  (left panel: 300 and 410 Myr. Right panel: 300 and 380 Myr). Dotted lines
  represent the magnitude and colour cuts adopted to select RC stars, whereas
  the dotdashed lines represent the magnitude cut, used to divide the two parts
  of the RC. Right panels: ratios between the stars in the upper portion of the
  RC and their total number as a function of the ratio of the young SSP (top
  panel: SSPs with ages of 300 and 410 Myr; bottom panel: SSPs with ages of 300
  Myr and 410 Myr). The ratio obtained in the observed DRC CMD is reported as a
  dashed line in both panels. The typical uncertainty ($\approx$ 10\%) in the
  calculated number ratio is also shown.} 
\label{f:countRC}
\end{figure*}

As stated in Section~\ref{s:mcsim}, we also simulated synthetic CMDs 
combining two SSPs with different ages. One set is obtained using SSPs with ages
of 300 Myr and 410 Myr and the other with ages of 300 Myr and 380 Myr. For both
cases, we derived a set of simulations in which we varied the ratio of the
number of stars in the young and the old SSP, from 10\% to 90\% in each of
them, with a step of 5\%. For each simulated CMD we obtained the corresponding 
pseudo-age distribution that we compared with the observed one.  

The fact that the stars from the two SSPs are mixed together in the
parallelogram used to derive the pseudo-age distribution, combined with our
purely statistical approach, cause an increase in the ``free parameters'' and
introduces a sort of degeneracy in the derived results. This is reflected in the
fact that we obtained more than one simulated pseudo-age distribution that
reasonably reproduces the observed one. These ``best-fitting'' pseudo-age
distributions are shown in Figure~\ref{f:cfr2}. In detail, the top panels show
the best fit achieved from the SSPs with ages of 300 Myr and 410 Myr (with mass
fractions of ``young'' stars of 40\% and 50\%, respectively), whereas in the
bottom panels we show the ones obtained from the SSPs with ages of 300 Myr and
380 Myr (with mass fractions of young stars of 40\% and 55\%, respectively) The
ages of the two SSPs and the ratio between them have been reported in each
panel. We note that the best-fit mass ratios of the young to the old population
seem to be around 1:1, consistent with the observed pseudo-age distribution,
where we see a hint of two peaks with a similar maximum value and hence a likely
number ratio of $\approx$\,1:1 for young vs.\ old stars in the cluster.   

To constrain which combination of two SSPs provides the best solution to
reproduce the observed CMD of NGC~1856, we performed two-sample
Kolmogorov-Smirnov (K-S) tests. The two-tailed {\it p} values are mentioned  in
Figure~\ref{f:cfr2} for each pseudo-age distribution. Taking these results at
face value, it seems that the combination of SSPs with ages of 300 and 380 Myr
provide a better solution with respect to the SSPs with ages of 300 and 410 Myr.
To further constrain the two-SSP fits, we pointed our attention to the
distribution of RC stars, adopting the following  approach. The left panels of
Figure~\ref{f:countRC} show the DRC CMDs zoomed in to the RC and the upper
portion of the MS, with superimposed isochrones from \citet{mari+08}, with the
same ages as those adopted in the SSPs (left sub-panel: 300 and 410 Myr; right
sub-panel: 300 and 380 Myr, respectively).  First, we selected cluster RC stars
by  applying colour and magnitude cuts (shown as dotted lines in the left panels
of Figure~\ref{f:countRC}). Then, we divide RC stars in two parts, applying a
magnitude threshold (shown as dotdashed lines in Figure~\ref{f:countRC})
coinciding with the brightest point of the RC of the ``old'' isochrone (i.e.,
F438W = 19.3 mag for the 410 Myr isochrone and F438W = 19.15 mag for the 380 Myr
isochrone). Finally, we counted the stars in the upper and lower portion of the
RC, both in the observed data and in each synthetic CMD obtained from the
associated Monte Carlo simulations.    

The ratios between the stars in the upper portion of the RC and their total
number as a function of the mass ratio of the young SSP in each simulation are
shown in the right panels of Figure~\ref{f:countRC} (for SSPs with ages of 300
Myr and 410 Myr in the top panel and for SSPs with ages of 300 Myr and 380 Myr
in the bottom panel, respectively). The ratios from the DRC CMD, obtained with
the different magnitude cuts, are shown as dashed lines. The typical uncertainty
on the mass ratios has been estimated to be of order 10\%, shown as an error bar
in both panels.  Taking these results at face value, it seems that the best
match is obtained from the synthetic CMD obtained from the SSPs with ages of 300
Myr and 380 Myr and with a fraction of young stars of 0.55\,$\pm$\,0.10 (cf.\
bottom right panel in Figure~\ref{f:cfr2}). 

Globally, the results presented in Section~\ref{s:ssp1} and \ref{s:ssp2} show
that a single SSP fails in reproducing the observed pseudo-age distribution
while a combination of two SSPs with different ages can provide a very good fit.
This   seems to indicate that a spread in age may be present in the cluster and
can not be categorically excluded as argued by \citet{bassil13}.  

\section{Constraints on ongoing star formation activity}
\label{s:excess}
\begin{figure}
\includegraphics[width=1\columnwidth]{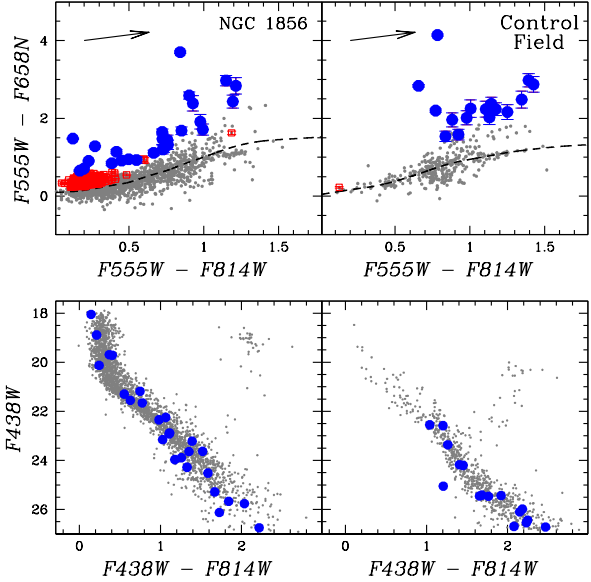}
\caption{Top panels: F555W - F814W vs F555W - F658N colour-colour diagram for
the stars inside the core radius (left panel) and in the control field (right
panel), representing the contamination from the background LMC stars (same field
adopted in Figure~\ref{f:cmdobs}). The dashed line represents the median  F555W
- F658N colour, representative of stars with no $H_{\alpha}$ excess. Stars with
a F555W - F658N colour at least $5 \sigma$ above that reference line, where
$\sigma$ is the uncertainty on the F555W - F658N colour of the star, are
reported as open red squares. Those among them with $EW_{H_{\alpha}} > 10 \AA$,
where the equivalent width of the $H_{\alpha}$ emission line ($EW_{H_{\alpha}}$)
is calculated using the Eq.~4 of \citet{dema+10} are reported as solid blue
circles.  Bottom Panels: CMDs of the cluster and of the control field with
overplotted stars selected from the F555W - F814W vs F555W - F658N colour-colour
diagram (solid blue circles).} 
\label{f:excess_ha}
\end{figure}

To study whether NGC~1856 \emph{presently} hosts ongoing star formation, we
performed a search for PMS stars by means of narrow-band imaging of the
H$\alpha$ line. H$\alpha$ emission is a good indicator of the PMS stage: the
presence of strong H${\alpha}$ emission ($EW_{H{\alpha}} > 10$ \AA) in young
stellar objects is normally interpreted as a signature of the mass accretion
process onto the surface of the object that requires the presence of an inner
disk \citep[see][and reference therein]{feimon99,whibas03}.  To do this, we used
the method described in \citet{dema+10}, which combines \emph{V} and \emph{I}
broad-band photometry with narrow-band $H{\alpha}$ imaging to identify all the
stars with excess $H{\alpha}$ emission and to measure their $H{\alpha}$
luminosity. 

In the top panels of Figure~\ref{f:excess_ha} we show the F555W\,$-$\,F814W vs.\
F555W\,$-$\,F658N colour-colour diagram for the stars inside the cluster core
radius (left top panel) and for the stars in the ``control field'', the same
field we used to derive the contamination by the background LMC population in
Figure~\ref{f:cmdobs} (right top panel). We used the median F555W\,$-$\,F658N
colour of stars with small ($<$ 0.05 mag) photometric uncertainties in each of
the three F555W, F814W, and F658N bands, as a function of F555W\,$-$\,F814W, to
define the reference template with respect to which  excess $H{\alpha}$ emission
is identified (shown by the dashed line in Figure~\ref{f:excess_ha}). We
selected a first candidate sample of stars with $H{\alpha}$ excess emission by
considering all those with a F555W\,$-$\,F658N color at least $5 \sigma$ above
that reference line, where $\sigma$ is the uncertainty on the F555W\,$-$\,F658N
colour of the star in question (shown as open red squares in
Figure~\ref{f:excess_ha}). Then we calculated the equivalent width of the
$H{\alpha}$ emission line ($EW_{H{\alpha}}$) from the measured color excess,
using the following equation from \citet{dema+10}: 
\begin{equation}
W_{eq}(H{\alpha}) = RW \times [1 - 10^{-0.4 \times (H{\alpha}-H{\alpha}^c)}]
\label{eq:ha_ex}
\end{equation}
where RW is the rectangular width of the filter (similar in definition to the
equivalent width of the line), which depends on the characteristic of the filter
\citep[for the adopted F658N filter, Rw = 17.68; see Table~4 in][]{dema+10}.
Finally, we considered as bona-fide candidate PMS stars those objects with
$EW_{H{\alpha}} > 10$ \AA\  \citep{whibas03}, shown as solid blue circles in
Figure~\ref{f:excess_ha}. The number of objects that show excess $H{\alpha}$
emission is very low and is roughly the same in both fields. In the bottom
panels of Figure~\ref{f:excess_ha}, we overplotted these objects (shown as solid
blue circles) on the F438W\,$-$\,F814W vs.\ F438W CMDs of the cluster and the
control field; the majority of them are located on the fainter part of the
CMD, where the photometric errors are larger and the contamination by background
LMC stars is higher. In fact the number of objects that show an $H{\alpha}$
excess is comparable within the errors in the two fields below the TO of the
background stellar population (i.e. F438W $\la$ 23 mag). Hence, the number and
the location of these objects in the CMDs suggest that they can be considered
spurious detections. For what concerns the handful of objects at brighter
magnitudes (i.e. F438W $\ga$ 22 mag), where the photometric errors are smaller
and the contamination is lower, we hypothesize that these can be stars in a
binary system in which mass transfer between the primary and the secondary star
is occurring. We exclude the possibility that these objects are true PMS stars
due to the fact that, in this case, we would have observed a significant number
of these objects at fainter magnitudes (i.e. lower masses and hence longer PMS
lifetime).   

In this context, it is worth to note that our magnitude detection limit for stars showing $H{\alpha}$ excess is around F438W $\simeq$ 26.5 mag, which, at the cluster age, corresponds to stars with masses of $\simeq$ 0.8 \Msun. The PMS lifetime for such stars is of the order of $\sim$ 120 Myr, indicating that if star formation is occuring within the cluster, with our method we should be able to observe a significant number of PMS objects. 

Hence, these results indicate that the level of ongoing star formation activity
is negligible in NGC~1856. In fact, the lack of low-mass stars showing an $H{\alpha}$ excess indicates that the star formation in the cluster must have stopped at least $\sim$ 120 Myr ago, in agreement with the conclusions derived from the pseudo-age distribution analysis.

\section{Constraints to the stellar rotation scenario in producing eMSTO's}
\label{s:rotation}

\begin{figure}
\includegraphics[width=1\columnwidth]{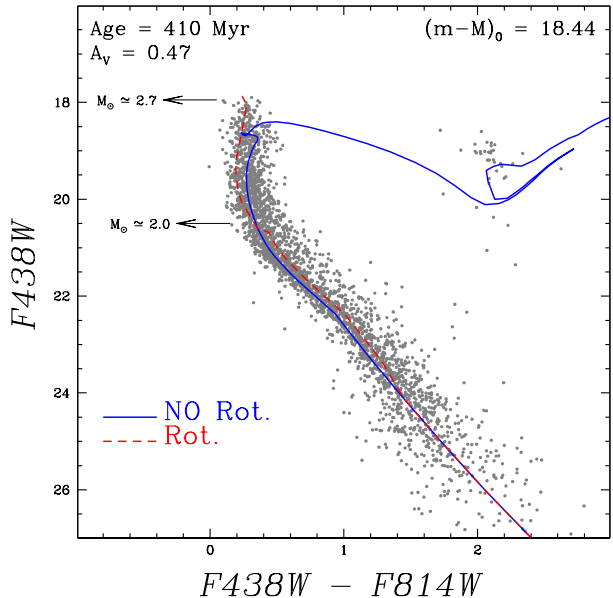}
\caption{Comparison between the non-rotating isochrone (blue solid line), from
  \citet{mari+08}, and its rotating counterpart (rec dashed line), derived as
  described in Section~\ref{s:rotation}. Isochrones are superposed on the
  cluster DRC CMD, along with adopted age, distance modulus $(m-M)_0$ and visual
  extinction $A_V$.} 
\label{f:cmdrot}
\end{figure}

In the literature, some alternative explanations that do not invoke the presence
of an extended star formation, have been proposed to explain the eMSTO
phenomenon in intermediate-age star clusters. In particular, the so called
``stellar rotation'' scenario \citep{basdem09,yang+13} suggest that eMSTOs can
be explained by a spread in rotation velocity among turn-off stars.
\citet{yang+13} calculated evolutionary tracks of non-rotating and rotating
stars for three different initial stellar rotation periods (approximately 0.2,
0.3 and 0.4 times the Keplerian rotation of ZAMS stars), and for two different
rotational mixing efficiencies (``normal'', $f_c = 0.03$ and ``enhanced'', $f_c
= 0.20$). From the isochrones, built from these tracks, they calculated the
widths of the MSTO region caused by stellar rotation as a function of cluster
age and translated them to age spreads. In particular, Figure~7 in
\citet{yang+13}  shows the equivalent width of the MSTO of star clusters caused
by rotation as a function of the cluster age, for different initial stellar
rotation periods and rotational mixing efficiencies. At the age of NGC~1856, all
the Yang et al.\ models with ``normal'' mixing efficiency (i.e.; $f_c = 0.03$)
indicate that rotation does not cause any appreciable spread.  On the other
hand, for their models with ``enhanced'' mixing efficiency (i.e.; $f_c = 0.20$),
the predictions vary from no spread for the model with the longest initial
stellar rotation period to a maximum of $\sim$ 100 Myr for the model with the
shortest period (i.e., the model for the highest rotation velocity). In this
case their model predicts a ``negative'' spread, in the sense that the rotating
model exhibits a bluer MSTO with respect to their non-rotating counterpart, thus
mimicking the presence of a younger population \citep[see top left panel of
Figure~6 in][]{yang+13}. 

To test the prediction of this rotating model in the specific case of NGC~1856,
we used the 410 Myr isochrone from \citet{mari+08} as a template for
non-rotating stars (i.e., the same isochrone as that plotted in
Figure~\ref{f:cmdcorr}), and we derived its rotating counterpart. To do so, we
calculated the shifts in colour and magnitude between the rotating and
non-rotating isochrones in Figure~6 of \citet{yang+13}, we then transformed
these shifts in our photometric system and we applied them to the
\citet{mari+08} isochrone. Figure~\ref{f:cmdrot} shows the DRC CMD of NGC~1856
with superimposed non-rotating and rotating isochrones (blue solid and red
dashed line, respectively). For the rotating isochrone, we plotted only the MS
part, since  the post-MS era of stellar evolution is not addressed by the
\citet{yang+13} models.  Overall, the rotating model plotted in
Figure~\ref{f:cmdrot} seems to reproduce the observed CMD reasonably well in the
upper portion of the MS, whereas the bottom portion of the MS is not as well
reproduced, the rotating model being somewhat too red with respect to the
observed MS. The latter is due to the fact that the rotating model bends toward
red colors (lower temperatures) for stellar masses ${\cal{M}} \la$ 2.0 \Msun.  

In conclusion, the rotating model of \citet{yang+13} that involves ``enhanced''
rotational mixing efficiency seems to provide a satisfactory fit to the observed
MS of NGC~1856, indicating that a spread in rotation velocity, beside an age
spread, could be a possible cause of the observed broadening of the MSTO
region.  However, we note that the predictions of the \citet{yang+13} models 
are \emph{not} consistent with the observations for intermediate-age ($\sim$
\,1--\,2 Gyr) star clusters featuring eMSTOs. This is \emph{especially the case
for the rotating models involving enhanced rotational mixing efficiency} (see
discussion in G14, in particular their Figure~7). It should however also be
recognized that the study of the creation of theoretical stellar tracks and
isochrones for rotating stars at various stages of stellar evolution, rotation
rates, and ages is still in relatively early stages. To date, no stellar
rotation velocity measurements have yet been undertaken in young and
intermediate-age star clusters in the Magellanic Clouds. We strongly encourage
such studies in the future in order to provide fundamental improvements of our
understanding of the possible relation between the eMSTO phenomenon and the
effects of stellar rotation.    

\section{Insights from Dynamical Analysis}
\label{s:dynamics}

As mentioned in the Introduction, a peculiar characteristic of eMSTOs in
intermediate-age star clusters in the Magellanic Clouds is that they are not
hosted by all such star clusters. To explain this phenomenon in  the context of
multiple stellar populations, G11a proposed a scenario in which eMSTOs can only
be hosted by clusters for which the escape velocity of the cluster is higher
than the wind velocities of ``polluter'' stars thought to provide the material
out of which the second stellar generation is formed, at the time such stars
were present in the cluster (we refer to this as ``the escape velocity
threshold'' scenario).  This scenario was developed further by G14 who studied
HST photometry of a sample of 18 intermediate-age (1\,--\,2 Gyr) star clusters
in the Magellanic Clouds that covered a variety of masses and core radii. They
found that all the clusters showing an eMSTO feature escape velocities $V_{\rm
esc} \ga$ 15 \kms\ out to ages of at least 100 Myr.  This age is equivalent to
the lifetime of stars of $\approx 5\;M_{\odot}$ \citep[e.g.,][]{mari+08}, and
hence old enough for the slow winds of massive binary stars and IM-AGB stars of
the first generation to produce significant amounts of ``polluted'' material out
of which second-generation stars may be formed.  Furthermore, C14 showed that
this threshold on $V_{\rm esc}$ is consistent with HST observations of four
low-mass intermediate-age star clusters ($\approx 10^4\; M_{\odot}$): the two
clusters that host eMSTOs have $V_{\rm esc} \ga$ 15 \kms\ out to ages of
$\sim$\,100 Myr, whereas  $V_{\rm esc} \la$ 12 \kms\ at all ages for the two
clusters \emph{not} exhibiting eMSTOs.  Hence, the critical escape velocity for
a star cluster to be able to retain the material ejected by first-generation 
polluter stars seems to be in the range of 12\,--\,15 \kms. This threshold is
consistent with wind velocities of IM-AGB stars and massive binary stars (see
G14 for a detailed discussion). Briefly, IM-AGB stars show  wind velocities in
the range 12\,--\,18 \kms \citep{vaswoo93,zijl+96}, whereas observed velocities
of ejecta of massive binary stars are in the range 15\,--\,50\footnote{We note that this measure have been derived from one system, RY Scuti, which is in a specific case of stable mass transfer, whereas most of the mass is ejected during the unstable phase. Lower velocities are expected during unstable mass transfer or the ejection of a common envelope.} \kms \citep{smit+02,smit+07}. 

With this in mind, following the results presented in the previous sections
where the analysis of pseudo-age distributions seems to suggest the presence of
multiple stellar populations in NGC~1856, we determined the structural
parameters and the dynamical properties of the cluster from our new HST/WFC3
data. 

\subsection{Structural parameters}
\label{s:kingmodel}
\begin{figure}
\hspace*{-0.5cm}
\includegraphics[width=0.7\columnwidth,angle=270]{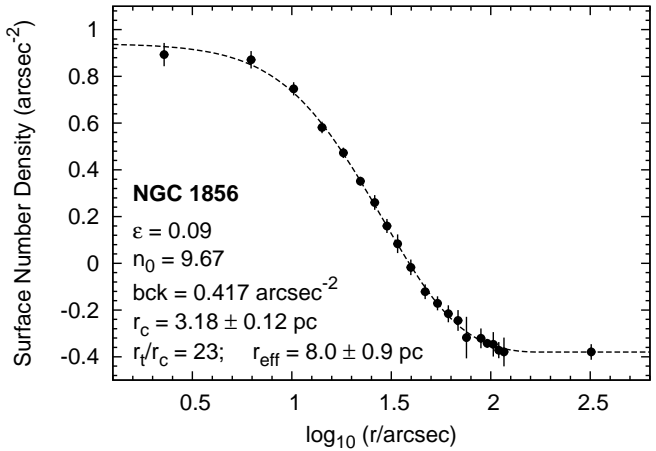}
\caption{Radial surface number density profile of NGC~1856. Black points
represent observed values. The dashed line represents the best-fit King model
(cf. equation~\ref{eq:King}) whose parameters are shown in the legend.
Ellipticity and effective radius $r_e$ are also shown in the legend. The radius
values have been converted to parsec by adopting the appropriate distance
modulus.} 
\label{f:king}
\end{figure}

We determined the radial surface number density distribution of stars, following
the procedure described in \citet{goud+09}. Briefly, we first determined the
cluster centre to be at reference coordinate ($x_c$, $y_c$) = (2986, 1767) with
an uncertainty of $\pm$ 5 pixels in either axis. In order to derive it, we first
created a two-dimensional histogram of the pixel coordinates adopting a bin size
of 50 $\times$ 50 pixels and then calculated the centre using a
two-dimensional Gaussian fit to an image constructed from the surface number
density values in the aforementioned two-dimensional histogram. This method
avoids  possible biases related to the presence of bright stars near the centre.
We derived the cluster ellipticity $\epsilon$ running the task {\it ellipse} 
within IRAF/STSDAS\footnote{STSDAS is a product of the Space Telescope Science
Institute, which is operated by AURA for NASA.} on the surface number density
images. Finally, we derived radial surface number densities by dividing the area
sampled by the images in a series of elliptical annuli, centered on the cluster,
and accounting for the spatial and photometric completeness in each annulus. 

\begin{figure}
\hspace*{+0.5cm}
\includegraphics[width=0.75\columnwidth]{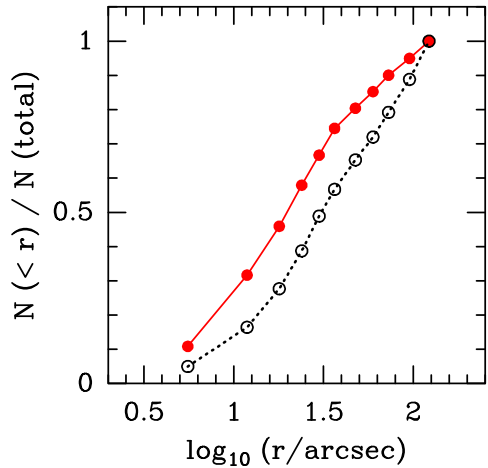}
\caption{Cumulative completeness-corrected radial distribution of bright versus
  faint stars. Red solid circles and solid line represent stars with F438W $<$
  20.5 mag while black open circles and dotted line represent stars with 21.0
  $<$ F438W $<$ 23.5 mag.}  
\label{f:mass_seg}
\end{figure}

We only considered stars brighter than the 30\% completeness limit in the core
region, corresponding to F438W $\leq$ 23.5 mag.  The outermost data point is
derived from the ACS parallel observations in a field located $\simeq$\,5\farcm5
from the cluster centre. The radial surface number density profile was fitted
using a \citet{king62} model combined with a constant background level,
described by the following equation: 
\begin{equation}
n(r) = n_0 \: \left( \frac{1}{\sqrt{1 + (r/r_c)^2}} - \frac{1}{\sqrt{1+c^2}}
\right)^2 \; + \; {\rm bkg} 
\label{eq:King}
\end{equation}
where $n_0$ is the central surface number density, $r_c$ is the core radius,  $c
\equiv r_t/r_c$ is the King concentration index ($r_t$ being the tidal radius),
and $r$ is the geometric mean radius of the ellipse ($r = a\,\sqrt{1-\epsilon}$,
where a is the semi-major axis of the ellipse). In Figure~\ref{f:king}, we show
the best-fit King model, obtained using a $\chi^2$ minimization routine. We
reported also the derived number density values along with other relevant
parameters. Our derived core radius of $r_c$ = 3.18 ($\pm$ 0.12) pc is
significantly larger than the literature value for NGC~1856 \citep[$r_c$ = 1.14
pc,][]{mclvan05}.  In reconciling this difference, we note that our King-model
fit was done using completeness-corrected surface number densities, whereas
\citet{mclvan05} used surface brightness data to derive the structural
parameters of the cluster. The latter method is sensitive to the presence of
mass segregation in the sense that mass-segregated clusters will appear to have
smaller radii when using surface brightness data than when using plain surface
number densities. To check whether NGC~1856 is actually mass segregated, we
derived the cumulative completeness-corrected radial distribution of bright and
faint stars  in the WFC3 image. ``Bright'' stars are selected by the magnitude
cut F438W $<$ 20.5 mag, whereas ``faint'' stars are selected in the magnitude
range 21.0 $<$ F438W $<$ 23.5 mag. The obtained cumulative radial distributions
are shown in Figure~\ref{f:mass_seg}. Note that the bright stars are clearly
more centrally concentrated than the faint ones, confirming that a significant
degree of mass segregation is present in the cluster.  

\subsection{Dynamical evolution and cluster escape velocity}
\label{s:dynevol}

\begin{table*}
\begin{center}
\begin{tabular}{cccccccc}
\hline
\hline
 V & Aper. & Aper. corr. & [Z/H] & $A_V$ & Age & $r_c$ & $r_e$ \\
 (1) & (2) & (3) & (4) & (5) & (6) & (7) &  (8) \\
\hline
 10.06 $\pm$ 0.15 & 31 & 0.38 & $-$0.30 & 0.47 & 300 & 3.18 $\pm$ 0.12 & 8.00 $\pm$ 0.90 \\
\hline
\end{tabular}
\caption{Physical properties of the star cluster. Columns (1): Integrated-light
$V$ magnitude from \citet{bica+96}. (2): Aperture radius in arcsec used for the
integrated-light  measurement. (3): Aperture correction in mag. (4): Metallicity
in dex. (5): Visual extinction in magnitude. (6): Mean age in Myr. (7): Core
radius in pc. (8): Effective radius in pc.}
\label{t:parameters}
\end{center}
\end{table*} 

\begin{table*}
\begin{center}
\begin{tabular}{|ccc|cccc}
\hline
\hline
 \multicolumn{3}{|c|}{log (${\cal{M}}_{\rm cl}/M_{\odot}$)} & 
 \multicolumn{4}{|c}{$V_{\rm esc}$ (\kms)}\\
Current & 10$^7$ yr w/o m.s. & 10$^7$ yr with m.s. &  Current & 
 10$^7$ yr w/o m.s. & 10$^7$ yr with m.s. &  10$^7$ yr ``plausible''\\
(1) & (2) & (3) & (4) & (5) & (6) & (7)\\
\hline
 5.07 $\pm$ 0.07 & 5.15 $\pm$ 0.10 & 5.37 $\pm$ 0.10 & 11.4 $\pm$ 0.7 & 
 13.7 $\pm$ 0.7 & 20.0 $\pm$ 0.7 & 17.1 $\pm$ 0.8 \\ 
\hline
\end{tabular}
\caption{Dynamical properties of the star cluster. Columns (1): Logarithm of
the adopted current cluster mass. (2-3): Logarithm of the adopted cluster mass
at an age of 10$^7$ yr without(with) the inclusion of initial mass segregation.
(4): Current cluster escape velocity at the core radius. (5-6): Cluster escape
velocity at the core radius at an age of 10$^7$ yr without(with) the inclusion
of initial mass segregation. (8) ``Plausible'' cluster escape velocity at an age
of 10$^7$ yr.}
\label{t:dynamics}
\end{center}
\end{table*}

We estimated the cluster mass and escape velocity as a function of time, going
back to an age of 10 Myr, after the cluster has survived the era of violent
relaxation and when the most massive stars of the first generation, proposed to
be candidate polluters in literature (i.e., FRMS and massive binary stars), are
expected to start losing significant amounts of mass through slow winds. The
current mass of NGC~1856 was determined from its integrated-light \emph{V}-band
magnitude listed in Table~\ref{t:parameters}. We determined the aperture
correction for this magnitude from the best-fit King model derived in Sect.\
\ref{s:kingmodel} by calculating the fraction of total cluster light encompassed
by the measurement aperture. After correcting the integrated-light \emph{V}
magnitude for the missing cluster light beyond the measurement aperture, we
calculated the total cluster mass adopting the values of $A_V, \mbox{[Z/H]}$,
and age listed in Table~\ref{t:parameters}. This was done by interpolation
between the ${\cal{M}}/L_V$ values in the SSP model tables of \citet{bc03},
assuming a \citet{salp55} initial mass function. The latter models were recently
found to provide the best fit (among popular SSP models) to observed
integrated-light photometry of LMC clusters with ages and metallicities measured
from CMDs and spectroscopy of individual stars in the  0.2\,--\,1 Gyr age range
\citep{pess+08}. 

We calculated the dynamical evolution of the star cluster following the
prescriptions of G14. Briefly, the evolution of cluster mass and radius was
evaluated with and without initial mass segregation, given the fundamental role
that the latter property plays in terms of the early evolution of the
cluster's expansion and mass loss rate \citep[e.g.,][]{mack+08b,vesp+09}. For 
the case of a model cluster with initial mass segregation, we adopted the
results of the simulation called SG-R1 in \citet{derc+08}, which involves a
tidally limited model cluster that features a level of initial mass segregation
of $r_e/r_{e,>1}$ = 1.5, where $r_{e,>1}$ is the effective radius of the cluster
for stars with ${\cal{M}} >$ 1 \Msun (see G14 for a detailed description of the
reasons for this choice).  

Table~\ref{t:dynamics} lists the derived masses and escape velocities at an age
of 10 Myr, obtained with and without initial mass segregation. Escape velocities
are calculated from the reduced gravitational potential $V_{\rm esc} (r,t) =
(2\Phi_{\rm tid} (t) - 2\Phi (r,t))^{1/2}$, at the core radius.  Here $\Phi_{\rm
tid}$ is the potential at the tidal (truncation) radius of the cluster. The
choice to calculate the escape velocity at the cluster core radius is related to
the prediction of the ``in situ'' scenario, where the second-generation stars
are formed in the innermost region of the cluster \citep{derc+08}.  For
convenience, we define $V_{\rm esc, 7} (r) \equiv V_{\rm esc}\,(r, t = 10^7 {\rm
yr})$, and refer to it as ``early escape velocity''. To estimate ``plausible''
values for $V_{\rm esc,\,7}$ we used a procedure that involves various results
from the compilation of Magellanic Cloud star cluster properties and N-body
simulations by \citet{mack+08b}. Briefly, they showed that the maximum core
radius seen among a large sample of Magellanic Cloud star clusters increases
approximately linearly with log(age) up to an age of $\sim$ 1.5 Gyr, namely from
$\simeq$ 2.0 pc at $\simeq$ 10 Myr to $\simeq$ 5.5 pc at $\simeq$ 1.5 Gyr.
Conversely, the {\it minimum} core radius is $\sim$ 1.5 pc throughout the age
range 10 Myr\,--\,2 Gyr. Using N-body modeling \citet{mack+08b} showed that this
behavior is consistent with adiabatic expansion of the cluster core in clusters
with different levels of initial mass segregation, in that clusters  with the
highest level of mass segregation experience the strongest core expansion (see
G14 for a more detailed discussion). Under this assumption, we derive that the
``plausible'' early escape velocity for NGC~1856 is  $V_{\rm  esc} = 17.1 \pm
0.8$ \kms, high enough to retain material shed by the slow winds of the
polluters stars.  

\begin{figure}
\includegraphics[width=0.8\columnwidth]{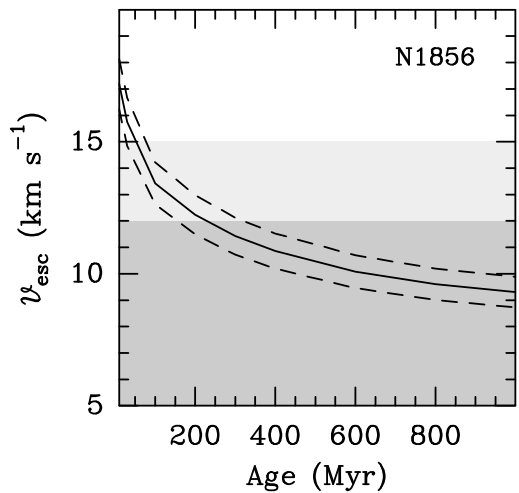}
\caption{Escape velocity $V_{\rm esc}$ as a function of time. The light grey
region represents the critical range of $V_{\rm esc}$ mentioned in Sect.\
\ref{s:dynamics}, i.e., 12\,--\,15 \kms. The region below 12  \kms, in which
$V_{\rm esc}$ is thought to be too low to permit retention of  material shed by
the first stellar generation, is shown in dark grey. $\pm 1 \sigma$ errors of
$V_{\rm esc}$ are shown by dashed lines.}   
\label{f:escvel}
\end{figure}

Figure~\ref{f:escvel} shows the escape velocity of NGC~1856 as a function of
age, derived based on the assumed level of mass segregation presented above. The
critical escape velocity range of 12\,--\,15 \kms\ derived by C14 and G14 is
depicted as the light grey region in Figure~\ref{f:escvel}. The region below 12
\kms, representing, as stated above, the velocity range in which eMSTOs are not
observed in intermediate-age star clusters (C14), is shown in dark grey.    We
note that $V_{\rm esc}$ for NGC~1856 is $\ga 14-15$ \kms\ out to an age of
$\approx$\,100 Myr. We recall that this is equivalent to the lifetime of stars
of $\approx 5\;M_{\odot}$ (i.e., IM-AGB stars, see \citealt{vendan09}) so that
the slow winds of massive binary stars and IM-AGB stars of the first generation
should have produced  significant amounts of ``polluted'' material within that
time. As mentioned above, this situation is similar to that shared among eMSTO
clusters of intermediate age in the Magellanic Clouds (G14). It thus seems
plausible that a significant fraction of that material may have been retained
within the potential well of NGC~1856, allowing second-generation stars to be
formed.  

Finally, we note that the mass and escape velocities of NGC~1856 were likely 
high enough for the cluster to be able to accrete a significant amount of
``pristine'' gas from its surroundings in the first several tens of Myr after
its birth. This would have constituted an additional source of gas to form
second-generation stars (\citealt{conspe11}; G14; but see also 
\citealt{basstra14}). 

\section{Summary and Conclusion}
\label{s:conclusion}

We presented results obtained from a study of new, deep HST/WFC3 images of the
young (age $\simeq$ 300 Myr) massive star cluster NGC~1856 in the LMC. After
correction for differential reddening, we compared the CMD of the cluster with
Monte Carlo simulations, both of one ``best-fit'' SSP and of  two SSPs of
different ages, in order to investigate the MSTO morphology and to quantify the
intrinsic width of the MSTO region of the cluster. We studied the physical and
dynamical properties of the cluster, deriving its radial surface number density
distribution and determining the evolution of cluster mass and escape velocity
from an age of 10 Myr to 1 Gyr, considering the possible effects of initial mass
segregation.  

The main results of the paper can be summarized as follows: 

\begin{itemize}
\item NGC~1856 shows a broad MSTO region whose width can not be explained by
  photometric uncertainties, LMC field star contamination, or differential
  reddening effects. Comparison with Monte Carlo simulations of a SSP shows that
  the observed pseudo-age distribution is significantly wider than that derived
  from a single-age simulation. Conversely, combining two SSPs with different
  ages, we obtain a set of pseudo-age distributions that reproduce the observed
  one quite well. By considering the luminosity function of the RC feature, we
  conclude that a best fit is achieved with a combination of SSPs with ages of
  300 Myr and 380  Myr and a mass fraction for the younger component of
  $\approx$\,55\%. The observed pseudo-age distribution shows two distinct peaks
  with a similar maximum value and an uniform decline towards younger and older
  ages, with respect to the peaks. However, the small separation in age between
  the two peaks prevents us to conclude whether the morphology of the MSTO can
  be better explained with a smooth spread in age or by two discrete bursts of
  star formation. These results do not agree with the conclusions of
  \citet{bassil13} who conclude that the CMD of NGC~1856 is consistent with a
  single age to within 35 Myr. We expect that this difference is due to the fact
  that our data are significantly deeper, namely $\sim$ 6 mag. Consequently, we
  suggest that the arguments of \citet{bassil13} against the ``age spread''
  scenario should be considered with caution.   

\item We use $V$, $I$, and H$\alpha$ images to select and study candidate
  (``putative'') pre-MS stars in NGC~1856.  The numbers of ``putative'' pre-MS
  stars in the cluster field and the control (background) field are found to be
  similar, suggesting that the detections can be considered spurious and/or
  associated to residual noise. This indicates that star formation is not
  \emph{currently} ongoing in the cluster.  

\item The ``stellar rotation'' scenario for the nature of the eMSTO phenomenon
  has been tested by evaluating rotating and non-rotating isochrones of the same
  age with the DRC CMD.  Overall, a reasonable range of rotation velocities
  seems to be able to reproduce the MSTO properties quite well, \emph{albeit
  only if the rotational mixing efficiency is significantly higher than
  typically assumed values}.  However, several properties of MSTOs and RCs among
  eMSTO star clusters in the age range of 1\,--\,2 Gyr are inconsistent with
  such high rotational mixing efficiencies. This seems to indicate that a range
  of stellar ages provides a more likely explanation of the eMSTO phenomenon
  (including the wide MSTO of NGC~1856) than does a range of stellar rotation
  velocities, altough, with the current data alone, the latter hypothesis cannot
  formally be discarded. In this context, new stellar rotation measurements in
  young and intermediate-age star clusters, combined with new models of
  theoretical tracks and isochrones for rotating stars will be of fundamental
  importance to address the role of stellar rotation in the eMSTO
  phenomenon.    
  
\item The dynamical properties of NGC 1856 derived from our data suggest that
  the cluster has an early escape velocity of $\approx$\,17 \kms, high enough to
  permit the retention of material shed by the slow winds of polluter stars
  (IM-AGB stars and massive binary stars). The cluster escape velocity  remains
  above the threshold value of 14\,--\,15 \kms\ for $\approx$ 80\,--\,100 Myr,
  long enough for the slow winds of IM-AGB  stars to have ejected their
  envelopes. This material would likely have been  available for
  second-generation star formation which could have caused the wide MSTO.
  Furthermore, these early escape velocities are consistent with observed wind
  velocities of ejecta of IM-AGB stars and with those seen in massive
  binary star systems.     
\end{itemize}

\section*{Acknowledgments}
Support for this project was provided by NASA through grant HST-GO-13011 from
the Space Telescope Science Institute, which is operated by the Association of 
Universities for Research in Astronomy, Inc., under NASA contract NAS5--26555.  
We made significant use of the SAO/NASA Astrophysics Data System during this
project. THP acknowledges support through FONDECYT Regular Project Grant No. 1121005 and BASAL Center for Astrophysics and Associated Technologies (PFB-06).

\newpage

\appendix

\section{Photometric Catalog}
\label{s:photometry}

In Table~\ref{t:phot} we report positions and magnitudes, with the associated errors, for the first ten objects in our final photometric catalog. The complete table of stellar photometry is available upon request.

\begin{table*}
\begin{center}
\begin{tabular}{c|cc|cccccccc}
\hline
\hline
ID & X & Y & F438W & err & F555W & err & F656N & err & F814W & err\\
 (1) & (2) & (3) & (4) & (5) & (6) & (7) & (8) & (9) & (10) & (11)\\
\hline
1 & 1277.992 & 16.072 &  27.066 & 0.280 & 99 & 99 & 99 & 99 & 23.549 & 0.195\\	
2 & 1279.163 & 19.076 & 99 & 99 & 27.625 & 0.262 & 22.958 & 0.317 & 99 & 99\\
3 & 1274.036 & 20.648 &  26.536 & 0.156 & 99 & 99 & 22.620 & 0.292 & 99 & 99\\
4 & 1300.449 & 22.857 &  27.221 & 0.289 & 99 & 99 & 22.664 & 0.239 & 99 & 99\\
5 & 1319.126 & 26.275 &  21.564 & 0.019 & 99 & 99 & 20.680 & 0.072 & 99 & 99\\
6 & 1296.704 & 26.879 &  27.217 & 0.377 & 99 & 99 & 22.607 & 0.315 & 99 & 99\\
7 & 1306.224 & 27.537 &  27.096 & 0.208 & 99 & 99 & 22.816 & 0.218 & 99 & 99\\
8 & 1269.086 & 28.177 &  27.537 & 0.248 & 99 & 99 & 23.518 & 0.462 & 99 & 99\\
9 & 1290.509 & 29.335 &  25.467 & 0.076 & 24.877 & 0.058 & 99 & 99 & 23.678 & 0.041\\
10 & 1278.772 & 31.292 &  24.668 & 0.059 & 24.051 & 0.045 & 99 & 99 & 23.047 & 0.034\\
\hline
\end{tabular}
\caption{Stellar photometry. (1): Identification number. (2-3): X and Y coordinates in pixels. (4-11): magnitudes and associated errors.}
\label{t:phot}
\end{center}
\end{table*}

\label{lastpage}

\end{document}